\documentclass[12pt]{iopart}
\pdfoutput=1
\usepackage{iopams} %Uncomment if AMS fonts required
\usepackage{citesort}
\usepackage{subfigure}
\usepackage{stackrel}
\usepackage{mathrsfs}
\usepackage{graphicx}% Include figure files
\usepackage{bm}% bold math
\usepackage{textcomp}%for ref $\text pm$

\begin{document}
%\nocite{*}%uncommon to show all refs (including uncitated)
\title{Frustration of signed networks: How does it affect the thermodynamic properties of a system? }
\author{Jiayu Cao$^1$, Ying Fan$^2$, and Zengzu Di$^2$}
\address{$^1$ Department of Physics, Beijing Normal University.}
\address{$^2$ School of Systems Science, Beijing Normal University.}
\ead{zdi@bnu.edu.cn}

\vspace{10pt}
%\begin{indented}
%	\item[]\today
%\end{indented}

\begin{abstract}
Signed networks with positive and negative interaction are widely observed in the real systems. The negative links would induce frustration, then affect global properties of the system. Based on previous studies, frustration of signed networks is investigated and quantified. Frustrations of  $\pm J$ (Edwards-Anderson) Ising model with a concentration $p$ of negative bonds, constructed on different networks, such as triangular lattice, square lattice and random regular networks (RRN) with connectivity $k=6$ are estimated by theoretical and numerical approaches. Based on the quantitative measurement of frustration, its effects on phase transitions characterized by order parameter $q_{EA}$ are studied. The relationship of critical temperature $T_c$ with the quantified frustration $\mu$ is given by mean-field theory. It shows that $T_c$ decreases linearly with frustration $\mu$ . The theory is checked by numerical estimations, such as the Metropolis algorithm and Replica Symmetric Population Dynamics Algorithm. The numerical estimates are consistent well with the mean-field prediction.
\end{abstract}

\tableofcontents

\section{Introduction}

Networks are fundamental tools to depict the structure of complex systems. In a social network, nodes represent individuals, and bonds represent relationship or other interaction. While some global properties of such networks have been investigated at structural and topological level \cite{newman2011structure,albert2000topology,Milo824,vega2007complex,Newman8577}, the content (friendship-collaboration-trust or aversion-competition-mistrust) of relationships (bonds) is also important. So signed networks with positive and negative links have attracted much attention. Signed networks go beyond the simple connected-disconnected networks. Besides social systems, activatory and inhibitory in neural networks, mutualism and competition in ecological interactions, positive and negative feedback in engineering control, spin glass models, all can be characterized by signed networks. They are at the interface between network theory, complexity science, and statistical physics \cite{SN1-1,SN1-2_also_Label[2],SN1-4,SN1-5,SN1-6,SN1-7}. Then the study of such signed networks will have influence on social science, neuroscience, biology, and statistical physics.

\begin{figure}
	\centering
	\includegraphics[width=0.7\linewidth]{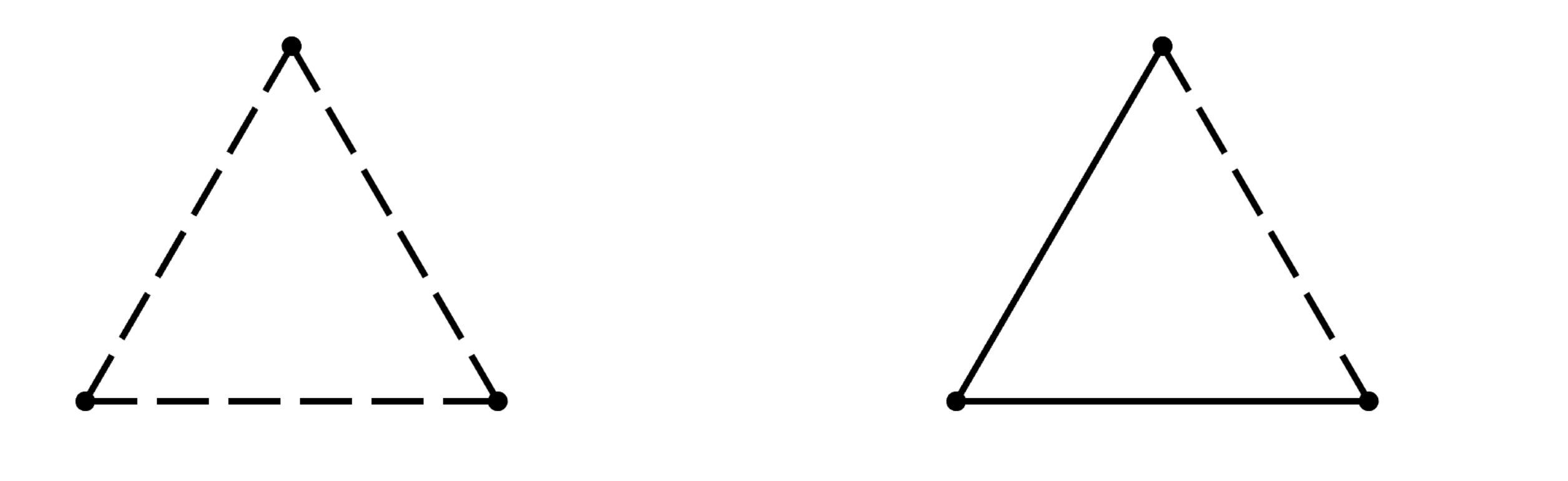}
	\caption{Examples of frustrated (unbalanced) relationships. Solid lines represent positive interaction. Dashed lines represent negative bonds.}
	\label{fig:frussn}
\end{figure}

\begin{figure} 
	\centering
	\includegraphics[width=0.7\linewidth]{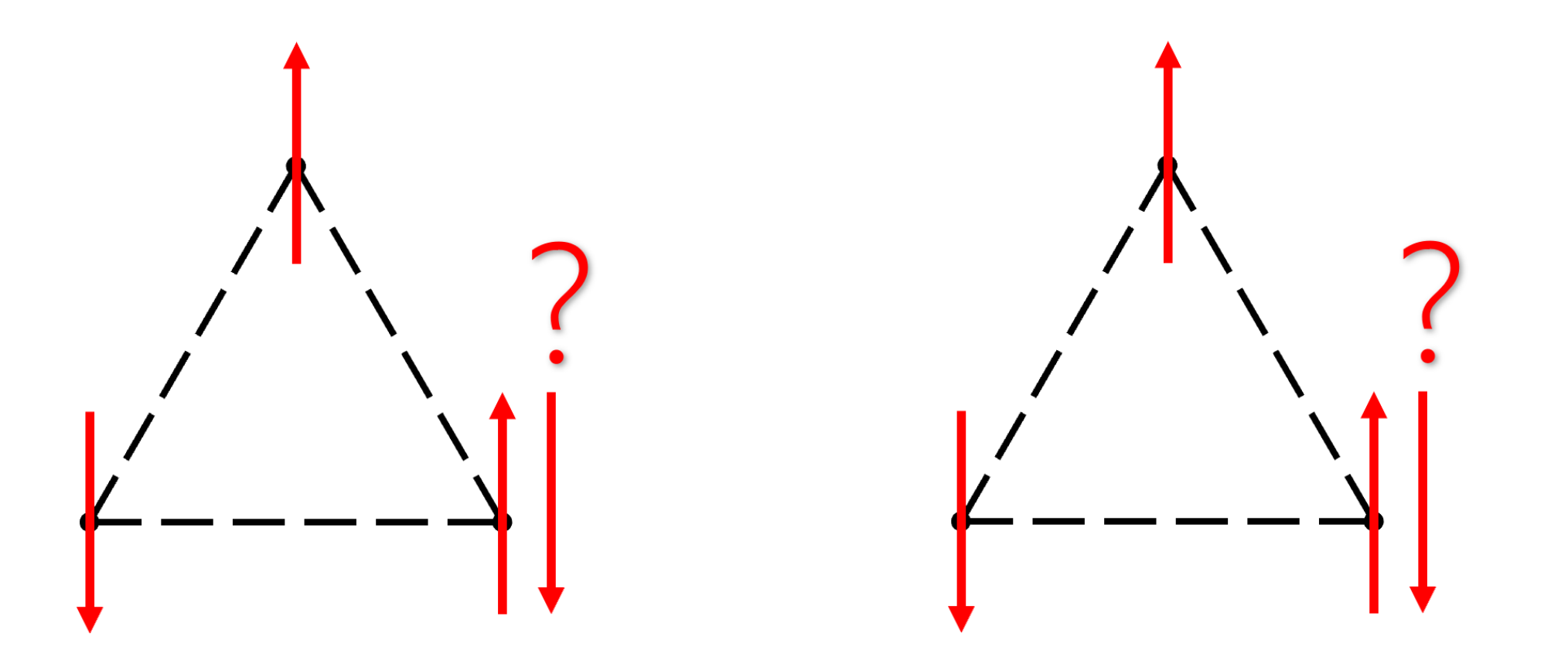}
	\caption{Examples of frustrated plaques. They are characterized by increased ground state energy ($-J$ in this example) and ground state degeneracy (6 in this example).}
	\label{fig:frus}
\end{figure}

The concept of frustration is important for the study of signed networks. A signed social network is frustrated when some unbalanced relationship such as ``the enemy of my enemy is my enemy" (and similar, equivalent statements) occur as showed in Fig.~\ref{fig:frussn}. Structural balance theory, formulated by Cartwright and Harary \cite{cartwright1956structural}, from Heider's work \cite{BalancedState}, affirms that signed social networks tends to be organized so as to avoid conflict situations mentioned above, corresponding to cycles of negative party. A balanced (organized) relationships, however, can also contain negative bonds. For example, ``an enemy's enemies are my friends" is a balanced relationship. From the Gauge theory\cite{TheoryoftheFrustration1[20]}, which we will discuss in detail latter, we can see that it is some effective negative bonds (they are also called frustrated edges) that are responsible for frustration effect. In the study of spin glasses, this effect is defined as topological constraints, which prevent neighboring spins from adopting a configuration with every bond energy minimized \cite{TheoryoftheFrustration1[20],TheoryofFrustration2}, and cause increasing of ground state energy and degeneracy, as showed in Fig.~\ref{fig:frus}. Frustration is responsible for many unexpected phenomena, and has attracted intensive attention \cite{FrusImportant1,diep2013frustration,lacroix2013introduction,PhysRevE2016AntiFrus}.

A lot of efforts are put to measure the distance from balanced states (strength of frustration) in the study of signed networks and spin glasses.
Definitions that are based on cycles and lines of structures are tried, such as relative portion of positive and negative ``cycles" and the proportion of bonds which must be removed or reversed to attain balance (see \cite{SG7} and \cite{SN3-DeterminingthedistanceAl,SN1-2_also_Label[2]}). Furthermore, increasing of ground state energy, usually called misfit parameter, is used to measure frustration in \cite{Kobe_frustration_11p_measure}.
More detailed analysis for $\pm J$ model on 2D lattice are also given in \cite{0_Asymptotic_Expansions,0_FrusGroundEngLoopEq}.
In fact, those approaches can be restated as finding effective negative edges, as we will see in Section \ref{Sec:MeasureFrus}, and by doing this we can see the intrinsic relationship between frustration and thermal properties clearly.
Frustration is also necessary to characterize spin glasses. Many studies have explored the thermodynamic effects of frustration:
The $J_{NN}$, $J_{NNN}$ frustration in \cite{SG8,SG9,SG10},
the positive-negative frustration of 3D Ising model in \cite{SG13,SG14},
2D anti-ferromagnetic triangular Ising spin lattice in \cite{SG15}, and
properties of 2D triangular $\pm J$ Ising spin glass in \cite{SG16}.

However, the quantitative relation between frustration and thermal properties is still an open problem. So it is interesting to show the effect of frustration on thermal properties, especially on phase transition based on the measurement of frustration. 
In Section~\ref{Sec:MeasureFrus}, we summarize the frustration in structure and energy aspect, including Gauge invariance, effective negative bonds, and its relation with ground state state energy. Then frustration of square and triangular $\pm J$ Ising model is given, based on the analysis of frustration of $L$ loops. Furthermore, the frustration of random regular network is given numerically both from the structure and energy estimation. In Section~\ref{Sec:ThermalStudy}, we go beyond the previous work, and get the relation between critical temperature and frustration. Specifically, taking $q_{EA}$ as the order parameter, we can get $T_c$ of a $\pm J$ Ising model with a fixed ratio of negative bonds $p$, given the critical temperature of the corresponding positive bonds Ising model.

\section{Measuring Frustration }\label{Sec:MeasureFrus}

In the study of disordered Ising system, the content of frustration is that topological
constraints prevent neighboring spins from adopting a configuration with every bond energy minimized \cite{TheoryoftheFrustration1[20]}.
In signed network study, some approaches of measuring frustration are developed (for a detail review see \cite{SG7}). For example, find the minimum number of edges to be switched or deleted or count the negative loops (loops have odd number of negative bonds). In spin glasses study, increasing of ground state energy is also used to measure frustration \cite{Kobe_frustration_11p_measure}. The quantitative definition of frustration will be discussed in the following sections based on the previous research. 

\subsection{Effective Negative Bonds}

Effective negative bonds is an important concept for measuring frustration. It is meaningful in both structural and physical aspect, and also gives us hints of phase transition especially single spin statistical dynamical behavior. The concept of effective negative bonds is intrinsically linked to the Gauge invariance, so we will introduce its structural meaning with help of Gauge invariance concept.

\subsubsection{Structural Meaning}
\begin{figure}
	\centering
	\includegraphics[width=0.8\linewidth]{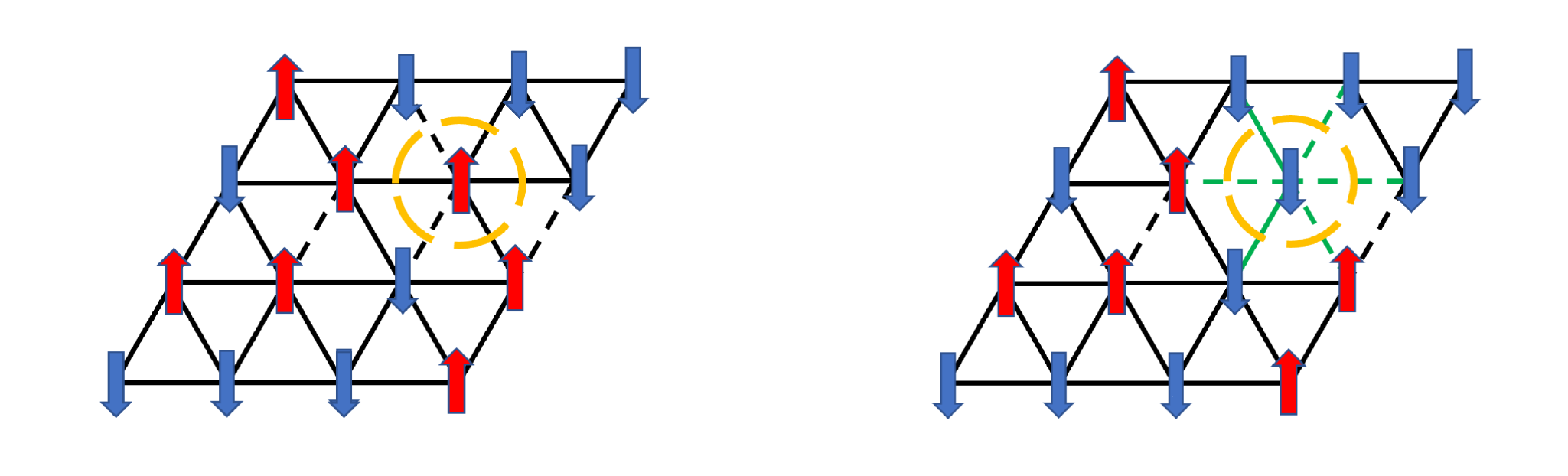}
	\caption{Illustration of a Gauge transformation between two models that are equivalent to each other. Given any micro-state of a model, there is one and only one corresponding micro-state for any of its equivalent models.}
	\label{fig:gaugetranspin}
\end{figure}
The Hamiltonian of a $\pm J$ Ising model, without external field, is
\begin{equation}\label{Hamiltonian}
E(\mathbf{s})=-\sum_{(i,j)} J_{i,j}s_i s_j/2 ,\qquad  \textrm{with} J_{i,j} =\pm J,
\end{equation}
which can be constructed on a network or lattice.

For a $\pm J$ Ising model with $B$ bonds, it has $2^B$ ways to get such a model, when the connection between nodes is fixed. However, different models with same connection may have the same statistical behavior of phase transition, such as Ising model and Mattis model \cite{EvenModel_SG1}. They have same $q_{EA}$ \cite{Theory_of_spin_glasses_P_W_Anderson}, and are actually in the same equivalent class, called switching class in the theory of signed graphs \cite{SignedGraphs}.
Gauge transformation \cite{TheoryofFrustration2,TheoryoftheFrustration1[20]}: 
\begin{equation}\label{eq:gauge_tran}
\sigma_i = \epsilon_i s_i,\qquad J^{\prime}_{ij}=\epsilon_i J_{ij}\epsilon_j, \quad \textrm{with} \quad \epsilon_i = \pm 1,
\end{equation}
can be carried between two states of two equivalent models in pair, see Fig.~\ref{fig:gaugetranspin} for an illustration.
This leaves the Hamiltonian and partition function invariant:
\begin{equation}\label{eq:gauge_invar}
H_J[\{s_i\}]=H_{J^{\prime}}[\{\sigma_i\}] \quad Z_J=Z_{J^{\prime}}.
\end{equation}
So models in an equivalence class should have some same thermal properties, and exhibit similar behaviors of phase transition.

\begin{figure}
	\centering
	\includegraphics[width=0.8\linewidth]{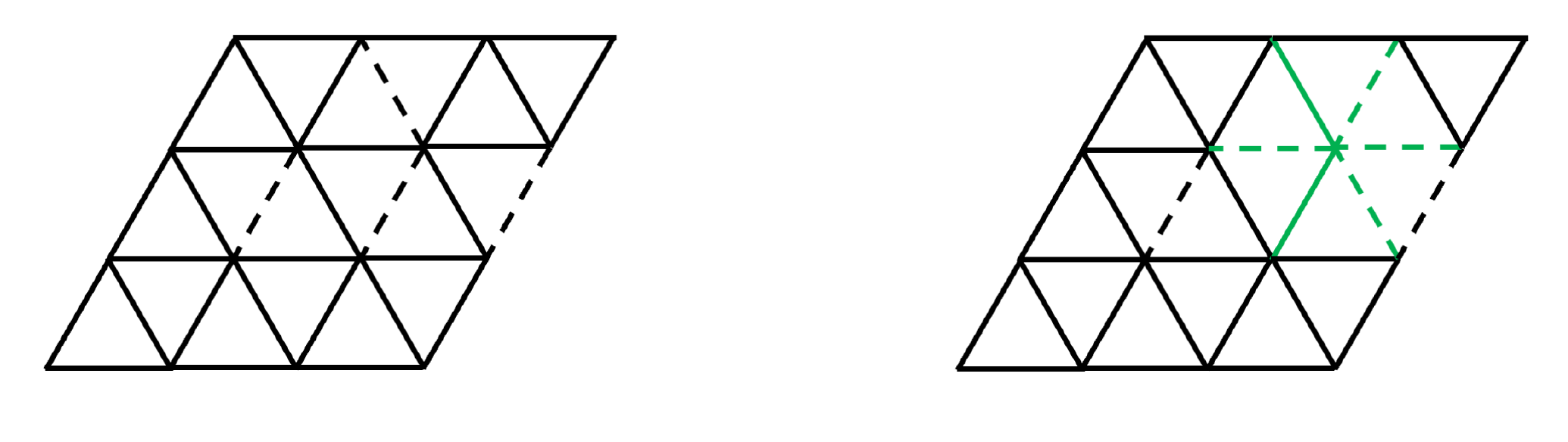}
	\caption{A way to get equivalent models. In fact the set of edges around one site is an subset of the cut set.}
	\label{fig:EqualModelAILLus}
\end{figure}
We can generate equivalent models of a model by flipping the sign of the edges through a cut set, which is a set of edges whose removal from a graph leaves the graph disconnected from structural aspect; see \cite{SN3-DeterminingthedistanceAl} for detailed proofs. For example, changing the sign of all bonds around one site is a simple way to get an equivalent model, as showed in Fig.~\ref{fig:EqualModelAILLus}, because the edges around a node is obviously a cut set. A famous example of such equivalences is Mattis model \cite{EvenModel_SG1}, which is equivalent to a pure Ising model.

Any one in an equivalent class is enough to generate all models in the whole class. For models in one class, there must be some models that have least negative bonds, and we call them ``effective negative bonds", because these bonds frustrate the model. By finding these models, we eliminate the apparent disorder. For instance, negative bonds of Mattis model don't bring frustration, and merely bring apparent disorder. We can see that only  effective negative bonds should affect some thermal-dynamic properties of the system, and the number of them $b$ is important for measuring strength of frustration. In fact, the portion of effective negative bonds can be used to measure the strength of frustration effect.

\subsubsection{Relation with Ground State Energy}

We will show that the ratio of effective negative bonds is 
actually half the relative change of ground state energy with respect to that of corresponding positive bonds Ising model.So the relative change of ground state energy is important for measuring frustration or distance from detailed balance states \cite{SN1-2_also_Label[2]}.

We can write the Hamiltonian and its Gauge transformation together:
\begin{equation}\label{eq:HamilMatGauge}
E(\mathbf{s})=-\mathbf{s}^T \mathscr{J}\mathbf{s}/2=-\sigma^T \mathscr{J}_{\epsilon} \sigma /2=E_{\epsilon}(\sigma).
\end{equation} 
Hamiltonian in (\ref{Hamiltonian}) is expressed in the first two terms of (\ref{eq:HamilMatGauge}), where $\mathbf{s}=[s_1,...,s_n]^T \in \mathbb{B}^n_2$; i.e. $s_i\in \{\pm1\},i=1,...n$, with $n$ equal to the number of nodes, and $ \mathscr{J}$ is an $n \times n$ symmetric matrix of entries $J_{i,j}=\pm J$.
$ \mathscr{J}$ actually describes the structure of a $\pm J$ Ising model with fixed connections, and is also named as adjacency matrix in signed networks literature.
Besides, using the same notations, the Gauge transformation (\ref{eq:gauge_tran}) is showed in last two terms of  (\ref{eq:HamilMatGauge}), where $\mathscr{J}_{\mathbf{\epsilon}}= T_{\mathbf{\epsilon}}\mathscr{J}T_{\mathbf{\epsilon}}$,
and $\sigma = T_\epsilon \mathbf{s} $; $T_\epsilon$ is a diagonal signature matrix $T_\epsilon=\textrm{diag}(\epsilon),\epsilon \in \mathbb{B}^n_2$, and in fact describes a Gauge transformation. An equivalent model described by $\mathscr{J}_{\epsilon}$ is generated in the transformation.

This transformation has the magic of expressing ground state energy in a structural way. If we can find an equivalent model, such that its  adjacency matrix  $\mathscr{J}_{\epsilon_0}$ has least $-J$ entries, then we can express ground state energy as:
\begin{equation}
E_0 = \min_{\mathbf{s} \in \mathbb{B}^n_2} {E(\mathbf{s})}
=	-\frac12 \mathbf{1}^T \mathscr{J}_{\epsilon_0}  \mathbf{1}.
\end{equation}
Effect of effective negative bonds becomes obvious, because ground states are attained when and only when all spins are aligned up or down for such equivalent models. The number of effective negative bond $b$ is also named $\delta$ in \cite{SN1-2_also_Label[2]} to measure the distance from detailed balance in social science study. Finally, the ground state energy can be expressed, in a structural way as:
\begin{equation}\label{eq:GE_Bonds}
E_0=(-B+2b)J.
\end{equation}
(\ref{eq:GE_Bonds}) shows that ground state energy and number of effective negative bonds has linear relationship. 
For convenience, we define
\begin{equation}\label{eq:DefineFrus}
\mu=2\frac{b}{B}=\frac{E_0^0-E_0}{E_0^0},
\end{equation}
where $E_0^0=-BJ$ is the ground state energy of models, of which negative bonds are replaced with positive ones. The parameter $\mu$ represents twice the proportion of effective negative bonds and the relative change of ground state energy with respect to a pure Ising model, and will be called frustration in the rest of this article. Some researchers also define this misfit parameter in energy aspect \cite{misfit}. Here we can see its clear structural meaning. The structural aspect of  effective negative bonds is crucial for study thermodynamic properties, as we will see in Section \ref{Sec:ThermalStudy}.

\subsection{Case Study: Square and Triangular Lattice, Random Regular Networks}\label{Sec:CaseStudy}

\subsubsection{Frustration of $L$ Loops}

The frustration of $\pm J$ $L$ loops ( loops with $L$ nodes and $L$ edges) will be measured first. The results are helpful for square and triangular lattice, and also for understanding frustration. A similar way of discussing frustration can be found in \cite{TheoryoftheFrustration1[20]}, which use curved squares to define frustration.
\begin{figure}
	\centering
	\includegraphics[width=0.7\linewidth]{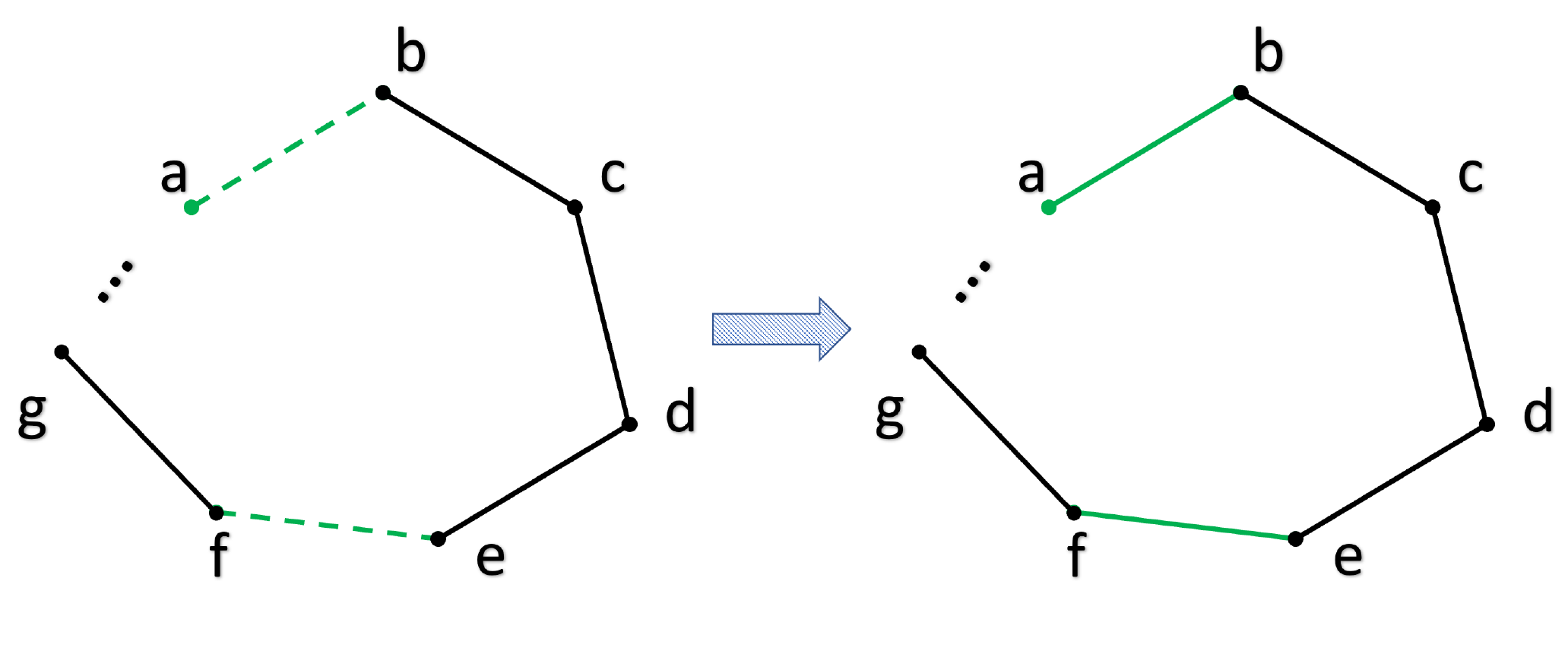}
	\caption{Flipping bond $ab$ and bond $fe$ at the same time, frustration of the loop doesn't change. In fact, any set has even number of edges in a loop is a cut set of this loop, because cutting any two edges makes the loop disconnected.}
	\label{fig:loopscutset}
\end{figure}

There are only two kinds of loops in frustration aspect: even loops (not frustrated) and odd loops (frustrated), because we can eliminate negative bonds in pair but maintain the frustration; see Fig.~\ref{fig:loopscutset}. Then, the corresponding frustration parameters $\mu$ for even loops and odd loops are:
\begin{equation}\label{eq:muL}
\mu_L^{even}=0,\quad \mu_L^{odd}=\frac{2}{L} .
\end{equation}

\begin{figure}
	\centering
	\includegraphics[width=0.7\linewidth]{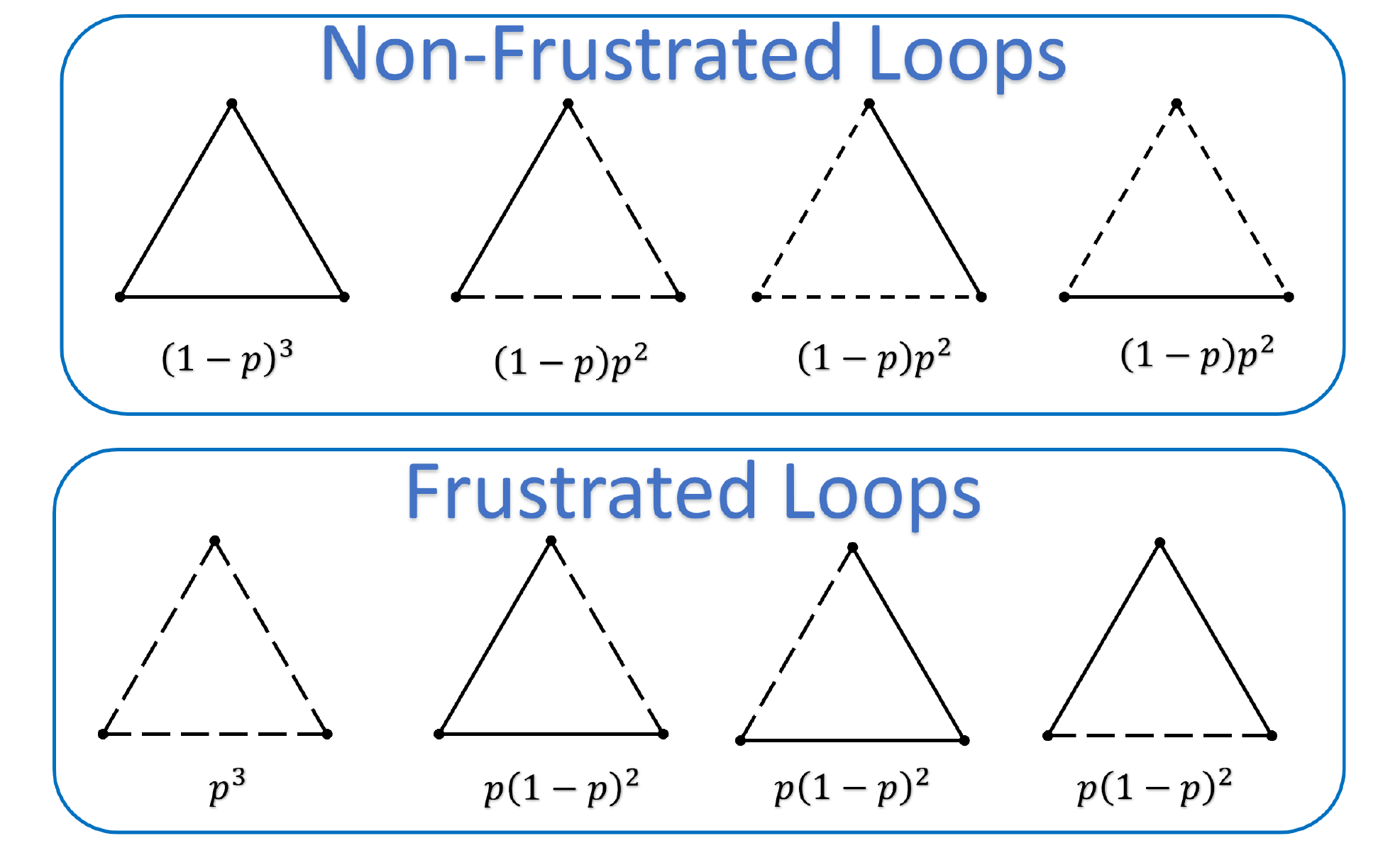}
	\caption{All possible loops and their possibilities for triangular plaques. $p$ is portion of negative bonds, or be the possibility of a bond to be negative in the thermodynamic limit.}
	\label{fig:triloopsfrusornot}
\end{figure}

For a $\pm J$ Ising model constructed on a loop, of which negative edges are assigned randomly with probability $p$, we can calculate the probability of getting even or odd loops. Listing all the possible loops and counting ratio of even and odd loops (an example is showed in Fig.~\ref{fig:triloopsfrusornot}), we can get possibility of getting an even or odd loop directly as 
$\sum_{k=0,2,4,...}^{k<=L} {L\choose k} p^k(1-p)^{L-k}$ and $\sum_{k=1,3,5,...}^{k<=L} {L\choose k} p^k(1-p)^{L-k}$ respectively. With help of combination equations, they can be rewritten as:
\begin{equation}\label{eq:weight}
p_L^{even}=\frac{1+(1-2p)^L}{2} ,\quad p_L^{odd}=\frac{1-(1-2p)^L}{2}.
\end{equation} 
Combining (\ref{eq:muL}) and (\ref{eq:weight}), we get expectation of frustration of $L$ loops:
\begin{equation}
\mu_L=\frac{1-(1-2p)^L}{L}.
\end{equation}

\subsubsection{Frustration of Square Lattice and Triangular Lattice}
Consider $\pm J$ square and triangular Ising models, where negative bonds are assigned randomly with probability $p$. In fact, both triangular and square lattices can be regard as combination of loops, even though these loops are not independent. However, frustration effect decrease at rate $ \frac{1}{L} $ with length of loops $L$, so it is reasonable to only consider the shortest loops in a given lattice. Then for square and triangular lattice, we have
\begin{equation}\label{eq:frustration SqTri}
\mu_{{square}}=\frac{1-(1-2p)^4}{4},\quad
\mu_{{triangular}}=\frac{1-(1-2p)^3}{3}\quad .
\end{equation}
This estimation is accurate when $p$ or $1-p$ is small. For the square lattice case, our formula, in fact, gives the lower bond on the frustration. However, the estimate is good enough for trend of frustration, because the upper bond and lower bond are close, see \cite{frustrationSQUpLowBonds}.

\begin{figure}%
	\centering 
	\subfigure[20*20 square lattice.]{ \includegraphics[width= 0.45 \linewidth]{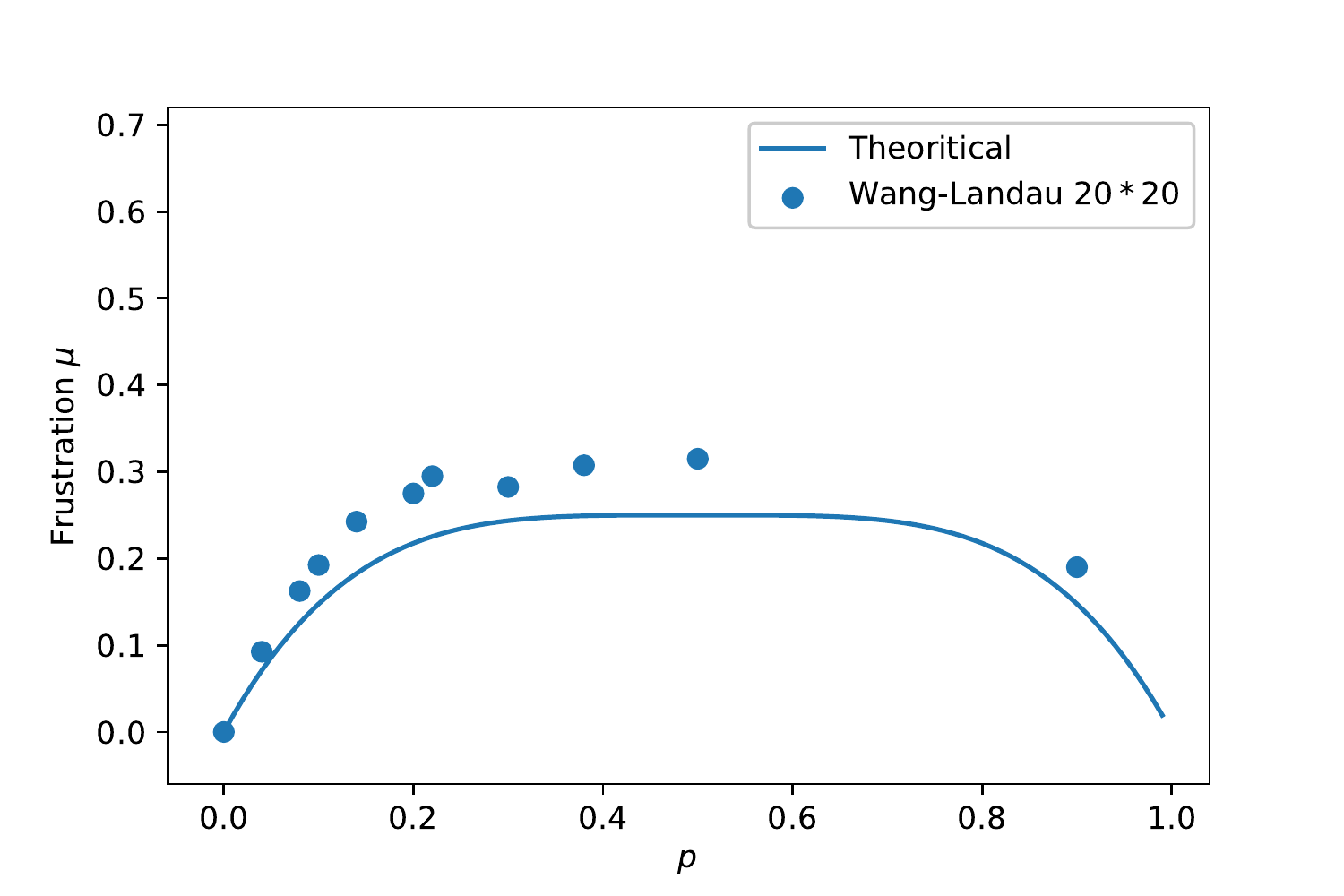}   }\qquad
	\subfigure[20*20 triangular lattice.]{ \includegraphics[width=0.45 \linewidth]{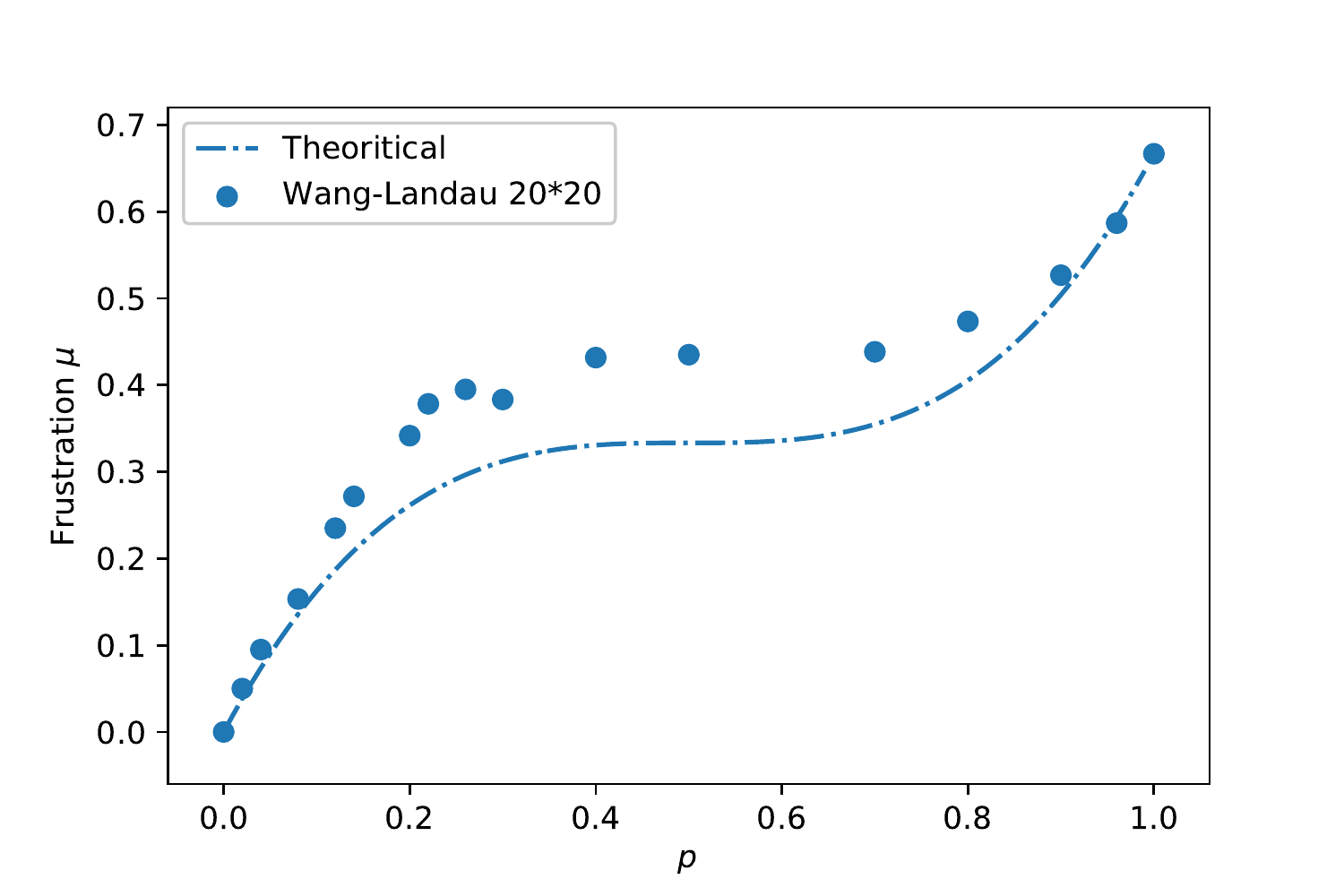}  }
	\caption{Use Wang-Landau algorithm to check estimation of frustration.}\label{fig:mulandau}
\end{figure}

To roughly check the estimated $\mu$, we use Wang-Landau (WL) algorithm \cite{Wanglandau_Algorithm} to calculate the ground state energy on 20*20 triangular and square lattices. We generate one sample for every different negative edge density $p$. According to \cite{EnergySizeEffect} , in which energy size effects are analyzed, WL on 20*20 lattices can give good estimation of $\mu$. As showed in Fig~.\ref{fig:mulandau}, the equations show reasonable trends and give fair estimations.

\subsubsection{$ \pm J $ Ising Model on Random Regular Networks}
\begin{figure}
	\centering
	\includegraphics[width=0.7\linewidth]{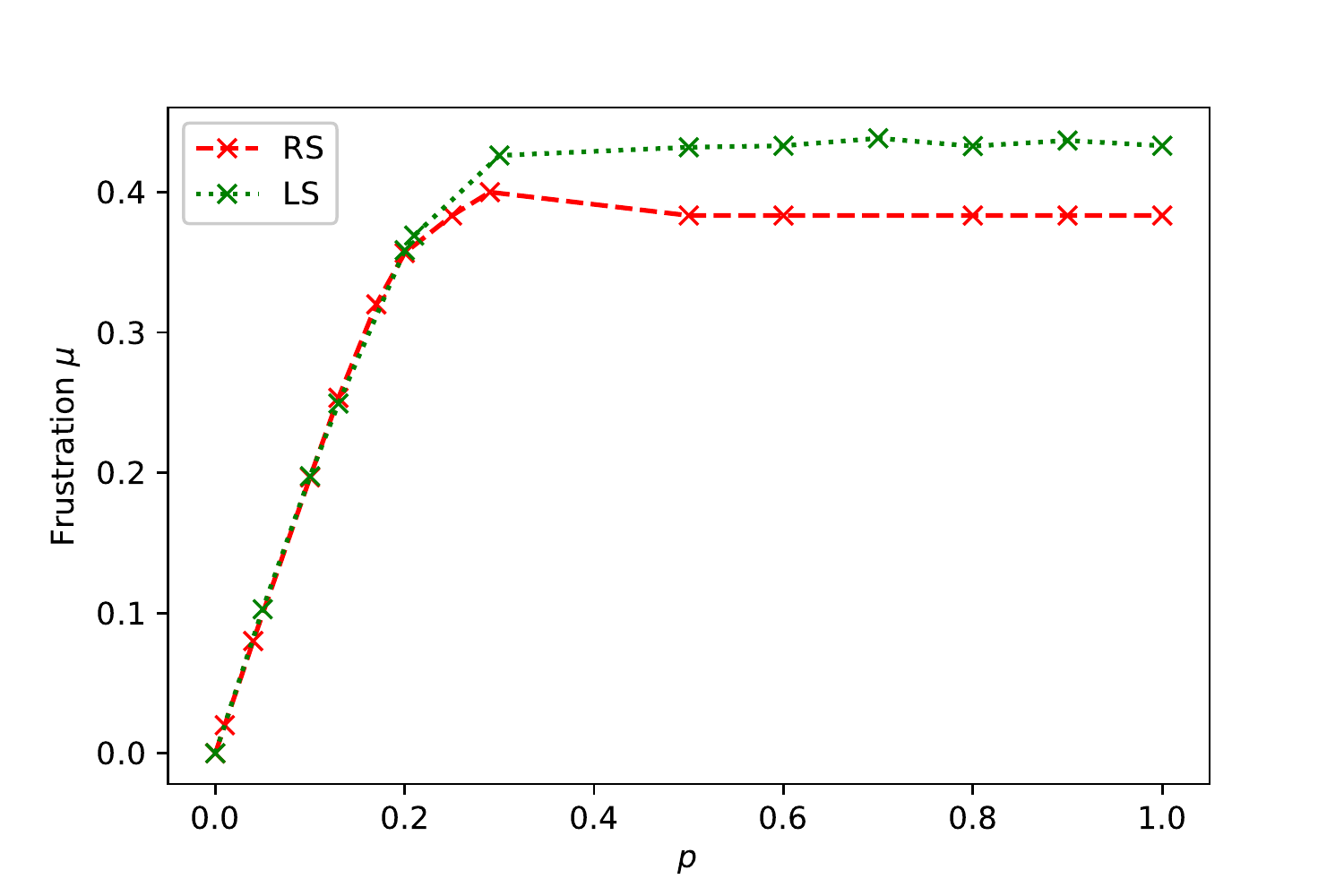}
	\caption{Frustration of $k=6$ random regular networks is estimated from topological structure and energy aspect.}
	\label{fig:PFrus6k}
\end{figure}

It is hard to get loops distribution of random regular Networks, so we only estimate the frustration $\mu$ numerically both in the structure and energy aspect. The results are showed in Fig.~\ref{fig:PFrus6k}.
In the structural way, a local search algorithm (LS) suggested in \cite{SN3-DeterminingthedistanceAl,SN1-2_also_Label[2]} is used to get the frustration. It in fact search for the matrix $\mathscr{J}_{\epsilon_0}$,  and gives up-bound of frustration. This algorithm can approach close to the frustration. Five random regular networks with $1000$ nodes are generated for every negative edge density $p$. For any sample of networks, the algorithm is tried three times to make up-bond lower.
We also get the frustration from its definition in energy aspect, using Replica Symmetric Population Dynamics (RS) \cite{PopulationDynamics} with the population of $10000$ to get ground state energy. Thermodynamic quantities including free energy density, mean energy density, Edwards-Anderson overlap order parameter, and entropy density are calculated with different temperature in RS. The whole population is updated $3000$ times, and thermodynamic quantities are calculated after $1000$ times updates. Mean energy density approaches ground state energy density when the entropy approaches zero. So we use this criterion to determine ground state energy.

\section{Thermodynamic Effects of Frustration 
}\label{Sec:ThermalStudy}

Some of researches have already studied the effects of frustration and negative bonds on thermodynamic properties, especially phase transition. Lots of phenomenological results are shown, but we still need to build a quantitative relationship based on the measurement of frustration. Here, we use the order parameter $q_{EA}$ suggested by Edwards and Anderson \cite{Theory_of_spin_glasses_P_W_Anderson} to depict the phase transition of $\pm J$ Ising models. This relation is helpful for understanding the phase transition on a particular structure and showing the relationship between structure and function of the complex systems. 

\subsection{Mean Field Approach}

Landau's mean field theory predicts critical temperature of a pure Ising model with $z$ nearest neighbors as $zJ$ (with $k_B = 1$ for convenience). We can generalize it to $z\langle J\rangle$ for disordered Ising model, where $\langle J\rangle$ is defined as the average of effective strength of interaction between spins. Then, for $\pm J$ Ising model, $\langle J\rangle$ is merely the average of effective positive and negative bonds, because the physical nature of phase transition is the competition between fluctuation and interactions. Therefore, the mean-field critical temperature of $\pm J$ Ising model can be expressed as
\begin{equation}\label{eq:J}
T_{c,m}=z<J>=zJ(1-\mu).
\end{equation}
Of course the mean field picture can not give the accurate critical temperature of pure and disordered Ising models. However, we assume the difference between critical temperature of a disordered Ising model and a pure Ising model ($T_c$ and $T_c^0$) is same as the difference between their mean-field ones ($T_{c,m}$ and $T_{c,m}^0$): 
\begin{equation}
T_c-T_c^0=T_{c,m}-T_{c,m}^0,
\end{equation}
Then by simple substitution, we can express the critical temperature as:
\begin{equation}\label{eq:T_c theory}
T_c=T_c^0-\mu T_{c,m}^0.
\end{equation}
Both mean-field and accurate critical temperature of pure Ising model ($T_c^0$ and $T_{c,m}^0$ ) can be found in previous research. Therefore, predicting critical temperature of a $\pm J$ Ising is possible, as long as we can get the frustration $\mu$. We have already get frustration $\mu$ of some models, so we can predict critical temperature of these models and compare it with numerical estimates.

\subsection{Numerical Results }

\begin{figure}%
	\centering 
	\subfigure[20*20 square lattice.]{ \includegraphics[width= 0.45 \linewidth]{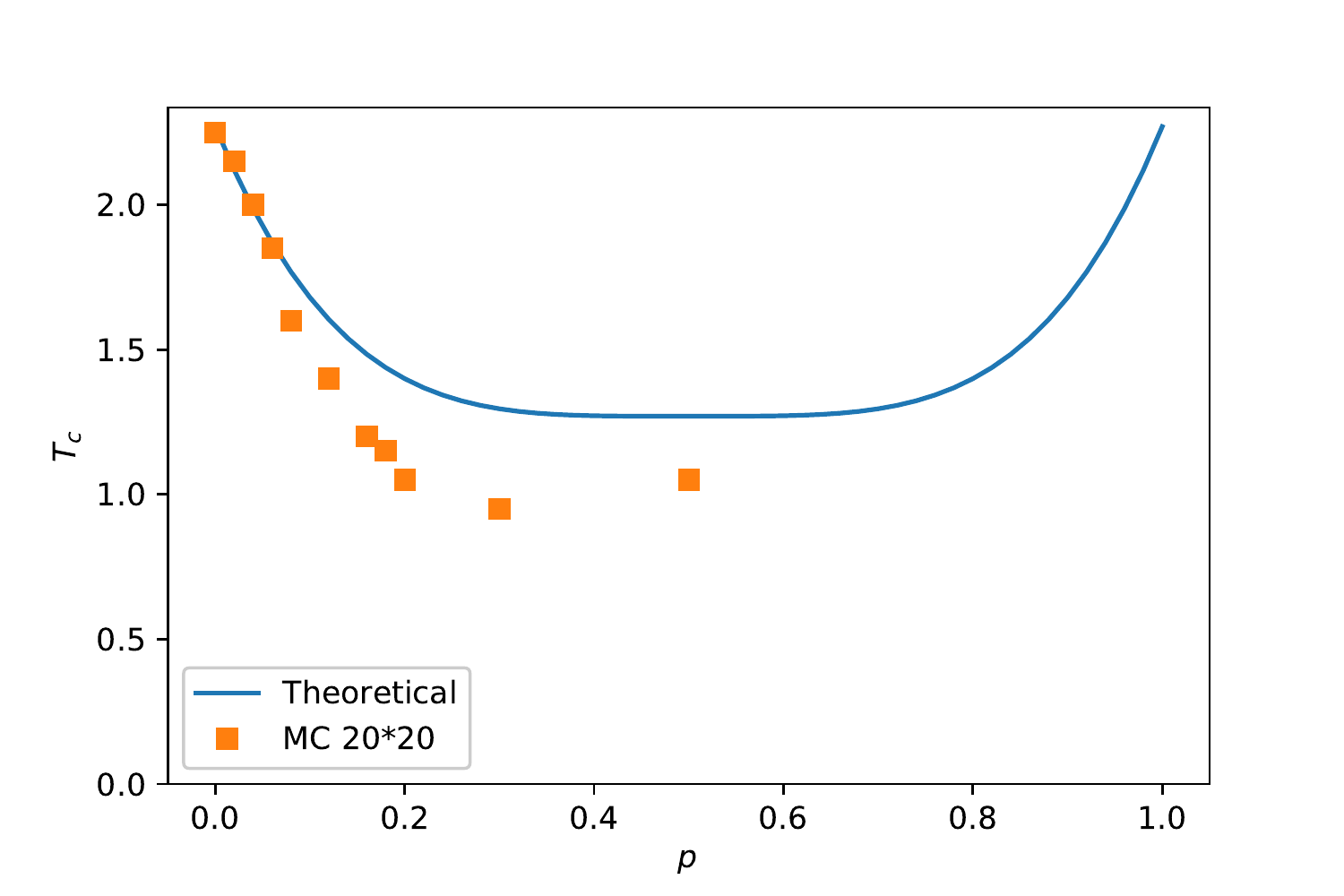}   }\qquad
	\subfigure[20*20 triangular lattice.]{ \includegraphics[width=0.45 \linewidth]{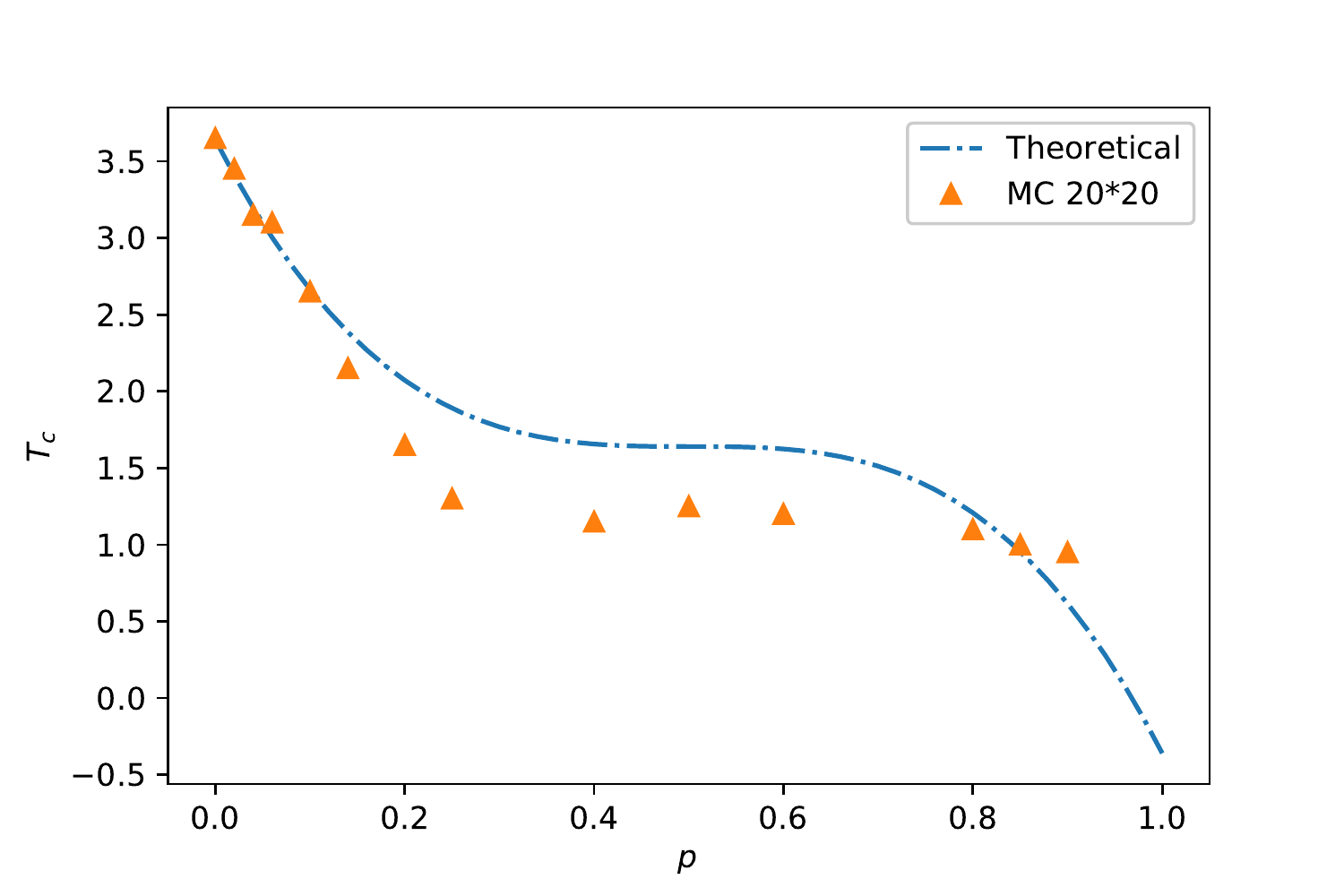}  }\\ 
	\subfigure[$k=6$, random regular networks.]{   \includegraphics[width=0.6 \linewidth]{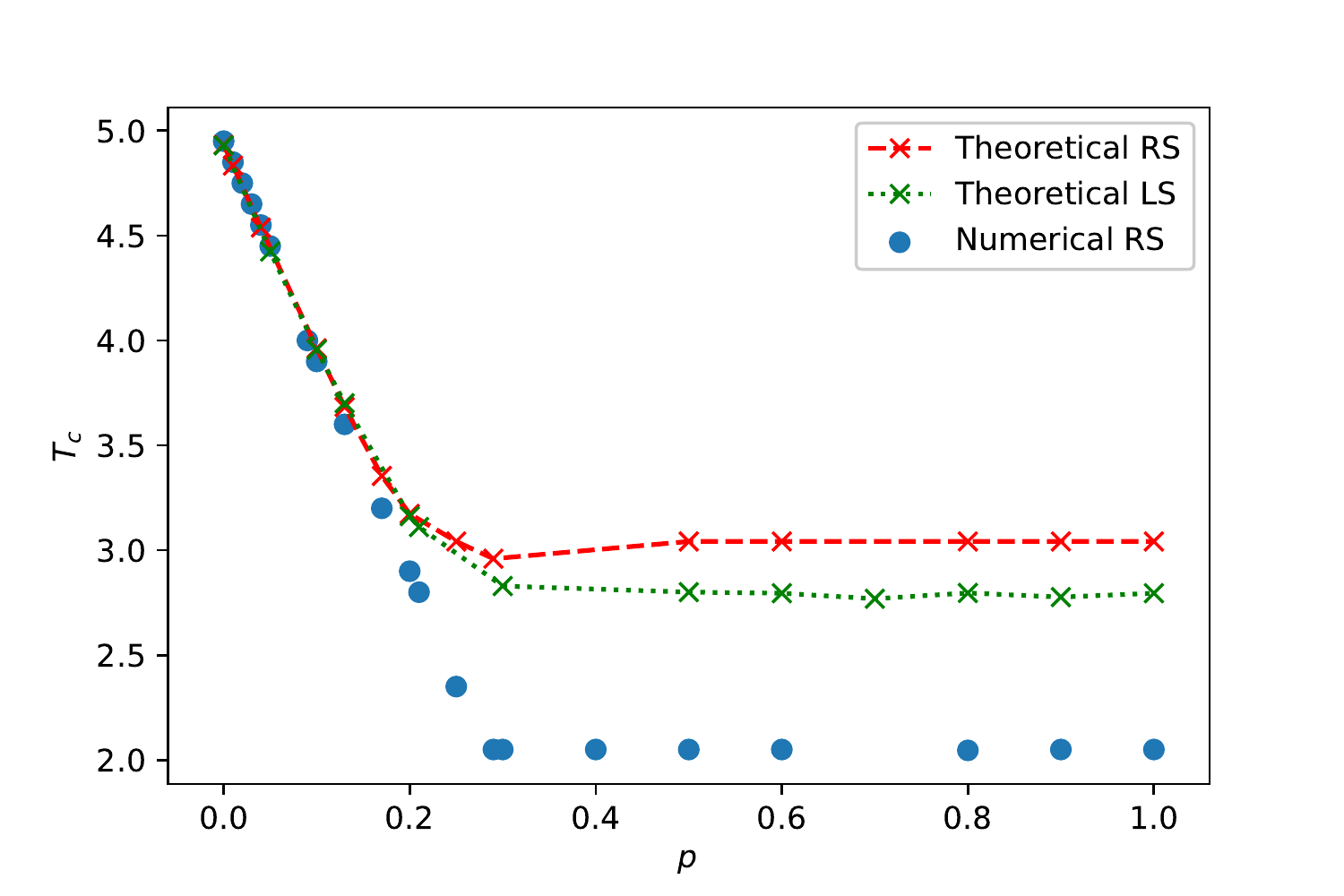}   }%
	\caption{Comparison between theoretically predicted (lines) and  estimated (symbols) critical temperature for three models as a function of $p$.}\label{fig:Tc3parra}
\end{figure}

\begin{figure}
	\centering
	\includegraphics[width=0.7\linewidth]{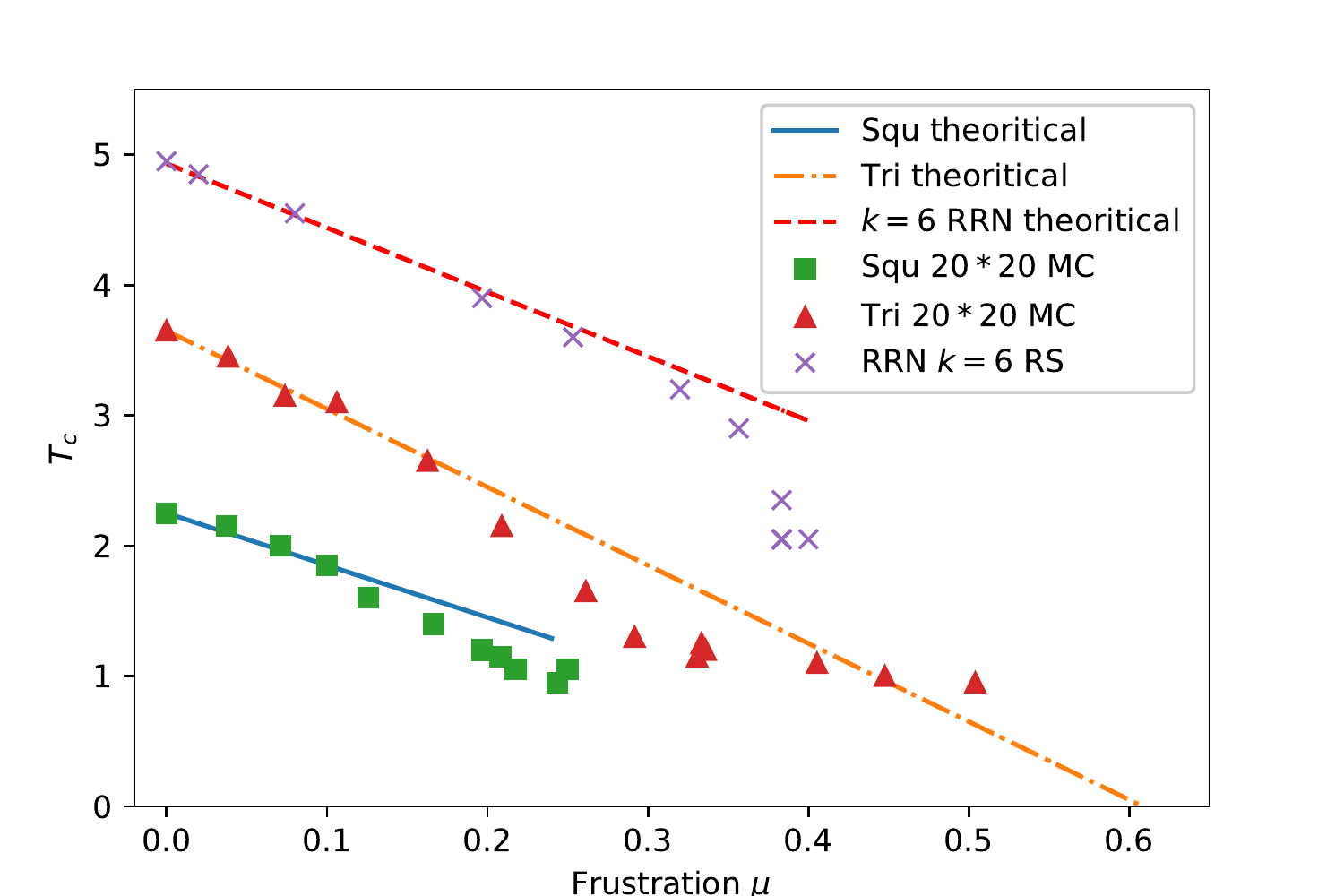}
	\caption{Comparison between theoretically predicted (lines) and  estimated (symbols) critical temperature for three models as a function of frustration $\mu$. }
	\label{fig:3inonesimutheo}
\end{figure}

Taking $J=1$ without losing any generalization, for square and triangular lattice, $T_{c,m}^0$ is $4$ and $6$ respectively. For a random regular networks $T^0_{c,m}$ is given by $(k-1)\tanh(\frac{1}{T^0_{c,m}})=1$.  $T_c^0$ for three graphs can be found on \cite{TcSqu,TcTri} and calculation using Replica Symmetric Mean Field theory. Here we take $T_c^0$ of square lattice as $2.27$, $T_c^0$ of triangular lattice as $3.64$, and $T_c^0$ of random regular networks with $k=6$ as $4.93$.
% Note that, we take $T^0_{c,m} = T_{c,m}$ for RRN, because this mean field theory can give good result for $p=0$ RRN. 
Besides, frustration $\mu$ of them has already been calculated theoretically or numerically in Section~\ref{Sec:CaseStudy}. Then, theoretical prediction line of critical temperature can be plotted by using (\ref{eq:T_c theory}).

Numerical estimation of critical temperature $T_c$ is done by Monte Carlo simulation. Note that determination of $T_c$ for a frustrated system is a tough problem. 
Furthermore, the trend of $T_c$ with change of frustration is what we are interested here. So only simulations on limited size of triangular and square lattices are carried. Metropolis algorithms (MC) are simulated on 20*20 square and triangular lattices with periodic boundary condition. $T_c$ is roughly estimated by the qualified change of $q_{EA}$ with the change of temperature. We use $2400000$ to $8800000$ Monte Carlo steps (MCS) to estimate the $q_{EA}$ for any temperature. Less MCS is required for pure Ising models to reach equilibrium sate. For the case of $8800000$ MCS, the system evolution was monitored after $8720000$ MCS to estimate $q_{EA}$ . More samples (up to $6$ samples) for temperatures near phase transition temperature are used to get average of $q_{EA}$. Also, step of $T$ near $T_c$ is $0.05$. $T_c$ for random regular networks with $k=6$ is got by Replica Symmetric Population Dynamic (RS) process \cite{PopulationDynamics}. The population of size $10000$ is updated $3000$ times totally, and in every update all cavity fields are updated. We begin to calculate $q_{EA}$ after $1000$ updates. This algorithm can give good results.

Theoretical and numerical results for relationship between proportion of negative bonds and $T_c$ are showed in Fig.~\ref{fig:Tc3parra}. The numerical results agree with our theory. Furthermore, by plotting relation between frustration and critical temperature in Fig.~\ref{fig:3inonesimutheo}, we could find that they really have linear relationship as suggested by mean field theory (\ref{eq:T_c theory}).

\section{Conclusion}

The behavior of a complex system is strongly affected by it underlying connectivity. Furthermore, besides the network topology, the sign of links, which describes the nature of interaction is also crucial for understanding the properties of a system. For signed networks, some negative loops may result in frustration. Actually, the frustration can be measured both from the structure and function aspects. Based on the previous researches on $\pm J$ Ising model, we quantify the frustration $\mu$ in energy and structure aspects. The frustration of $\pm J$ Ising model on square lattice, triangular lattice and random regular networks are given theoretically or by numerical estimation.

Furthermore, the thermodynamic effects of frustration are investigated. We put efforts to get relation between critical temperature and frustration for $\pm J$ Ising model. Assuming that mean field theory gives difference of critical temperature between a disordered and corresponding pure Ising model ($T_{c,m}-T_{c,m}^0 = T_c-T_c^0$), we give a formula to describe critical temperature of $\pm J$ Ising models. By estimating $T_c$ numerically, we show that the theory gives good estimation and picture of the phase transition for $\pm J$ Ising model. 

We have only hinted at the the effects of frustration on critical temperature in energy and structure aspects on frustrated signed networks. It is safe to assume that exploration of frustrated signed networks will reveal more unexpected phenomena. Detailed investigation on scaling properties of disordered Ising model, or other dynamical process on signed networks including random walks, epidemics, synchronization, transportation, and so on, could let us get more knowledge about the effects of frustration on complex systems. 

\section{Acknowledgments}
This project is supported by National Natural Science Foundation of China (NSFC) with Grants No. 71731002 and No. 61573065. J. Cao is supported by the Beijing Normal University Research Fund for Talented Undergraduates.
\section{References}
\bibliography{ref_alpha2}% Produces the bibliography via BibTeX.

\begin{thebibliography}{10}

\bibitem{newman2011structure}
Mark Newman, Albert-Laszlo Barabasi, and Duncan~J Watts.
\newblock {\em The structure and dynamics of networks}, volume~19.
\newblock Princeton University Press, 2011.

\bibitem{albert2000topology}
R{\'e}ka Albert and Albert-L{\'a}szl{\'o} Barab{\'a}si.
\newblock Topology of evolving networks: local events and universality.
\newblock {\em Physical review letters}, 85(24):5234, 2000.

\bibitem{Milo824}
R.~Milo, S.~Shen-Orr, S.~Itzkovitz, N.~Kashtan, D.~Chklovskii, and U.~Alon.
\newblock Network motifs: Simple building blocks of complex networks.
\newblock {\em Science}, 298(5594):824--827, 2002.

\bibitem{vega2007complex}
Fernando Vega-Redondo.
\newblock {\em Complex social networks}.
\newblock Number~44. Cambridge University Press, 2007.

\bibitem{Newman8577}
M.~E.~J. Newman.
\newblock Modularity and community structure in networks.
\newblock {\em Proceedings of the National Academy of Sciences},
  103(23):8577--8582, 2006.

\bibitem{SN1-1}
T.~Antal, P.~L. Krapivsky, and S.~Redner.
\newblock Dynamics of social balance on networks.
\newblock {\em Phys. Rev. E}, 72:036121, Sep 2005.

\bibitem{SN1-2_also_Label[2]}
Giuseppe Facchetti, Giovanni Iacono, and Claudio Altafini.
\newblock Computing global structural balance in large-scale signed social
  networks.
\newblock {\em Proceedings of the National Academy of Sciences},
  108(52):20953--20958, 2011.

\bibitem{SN1-4}
K.~Kulakowski.
\newblock Some recent attempts to simulate the heider balance problem.
\newblock {\em Computing in Science Engineering}, 9(4):80--85, July 2007.

\bibitem{SN1-5}
Seth~A. Marvel, Steven~H. Strogatz, and Jon~M. Kleinberg.
\newblock Energy landscape of social balance.
\newblock {\em Phys. Rev. Lett.}, 103:198701, Nov 2009.

\bibitem{SN1-6}
Aravind Srinivasan.
\newblock Local balancing influences global structure in social networks.
\newblock {\em Proceedings of the National Academy of Sciences},
  108(5):1751--1752, 2011.

\bibitem{SN1-7}
Michael Szell, Renaud Lambiotte, and Stefan Thurner.
\newblock Multirelational organization of large-scale social networks in an
  online world.
\newblock {\em Proceedings of the National Academy of Sciences},
  107(31):13636--13641, 2010.

\bibitem{cartwright1956structural}
Dorwin Cartwright and Frank Harary.
\newblock Structural balance: a generalization of heider's theory.
\newblock {\em Psychological review}, 63(5):277, 1956.

\bibitem{BalancedState}
Fritz Heider.
\newblock Attitudes and cognitive organization.
\newblock {\em The Journal of Psychology}, 21(1):107--112, 1946.
\newblock PMID: 21010780.

\bibitem{TheoryoftheFrustration1[20]}
G.~{Toulouse}.
\newblock {\em {Theory of the frustration effect in spin glasses: I}}, pages
  99--103.
\newblock World Scientific Press, November 1987.

\bibitem{TheoryofFrustration2}
J~Vannimenus and G~Toulouse.
\newblock Theory of the frustration effect. ii. ising spins on a square
  lattice.
\newblock {\em Journal of Physics C: Solid State Physics}, 10(18):L537, 1977.

\bibitem{FrusImportant1}
Leon Balents.
\newblock Spin liquids in frustrated magnets.
\newblock {\em Nature}, 464(7286):199, 2010.

\bibitem{diep2013frustration}
HT~Diep and H~Giacomini.
\newblock Frustration—exactly solved frustrated models.
\newblock In {\em Frustrated Spin Systems}, pages 1--58. World Scientific,
  2013.

\bibitem{lacroix2013introduction}
Claudine Lacroix, Philippe Mendels, and Fr{\'e}d{\'e}ric Mila.
\newblock {\em Introduction to Frustrated Magnetism: Materials, Experiments,
  Theory}.
\newblock Springer, 2013.

\bibitem{PhysRevE2016AntiFrus}
Unjong Yu.
\newblock Ising antiferromagnet on the 2-uniform lattices.
\newblock {\em Phys. Rev. E}, 94:022112, Aug 2016.

\bibitem{SG7}
Frank Harary.
\newblock On the measurement of structural balance.
\newblock {\em Behavioral Science}, 4(4):316--323.

\bibitem{SN3-DeterminingthedistanceAl}
G.~Iacono, F.~Ramezani, N.~Soranzo, and C.~Altafini.
\newblock Determining the distance to monotonicity of a biological network: a
  graph-theoretical approach.
\newblock {\em IET Systems Biology}, 4(3):223--235, May 2010.

\bibitem{Kobe_frustration_11p_measure}
S.~Kobe and T.~Klotz.
\newblock Frustration: How it can be measured.
\newblock {\em Phys. Rev. E}, 52:5660--5663, Nov 1995.

\bibitem{0_Asymptotic_Expansions}
E.E. Vogel and W.~Lebrecht.
\newblock Rapidly converging asymptotic expansions in {\textpm} j ising
  lattices.
\newblock {\em Zeitschrift f{\"u}r Physik B Condensed Matter}, 102(1):145--151,
  Mar 1996.

\bibitem{0_FrusGroundEngLoopEq}
W.~Lebrecht, E.E. Vogel, and J.F. Valdés.
\newblock Ising model on mixed two-dimensional lattices.
\newblock {\em Physica B: Condensed Matter}, 320(1):343 -- 347, 2002.
\newblock Proceedings of the Fifth Latin American Workshop on Magnetism,
  Magnetic Materials and their Applications.

\bibitem{SG8}
A~K Murtazaev, M~K Ramazanov, and F~A Kassan-Ogly.
\newblock Frustrations and phase transitions in the ising model on square
  lattice.
\newblock {\em Journal of Physics: Conference Series}, 510(1):012026, 2014.

\bibitem{SG9}
M.K. Ramazanov, A.K. Murtazaev, and M.A. Magomedov.
\newblock Thermodynamic, critical properties and phase transitions of the ising
  model on a square lattice with competing interactions.
\newblock {\em Solid State Communications}, 233:35 -- 40, 2016.

\bibitem{SG10}
A.K. Murtazaev, M.K. Ramazanov, and M.K. Badiev.
\newblock Critical properties of the two-dimensional ising model on a square
  lattice with competing interactions.
\newblock {\em Physica B: Condensed Matter}, 476:1 -- 5, 2015.

\bibitem{SG13}
Giacomo Ceccarelli, Andrea Pelissetto, and Ettore Vicari.
\newblock Ferromagnetic-glassy transitions in three-dimensional ising spin
  glasses.
\newblock {\em Phys. Rev. B}, 84:134202, Oct 2011.

\bibitem{SG14}
F.~Krzakala and O.~C. Martin.
\newblock Absence of an equilibrium ferromagnetic spin-glass phase in three
  dimensions.
\newblock {\em Phys. Rev. Lett.}, 89:267202, Dec 2002.

\bibitem{SG15}
Fometio~S Takengny, SS~Zekeng, Youmbi~B Sitamtze, F~Tchoffo, and E~Maga.
\newblock Frustration in 2d anti-ferromagnetic triangular ising spin lattice: a
  monte carlo study.
\newblock {\em The African Review of Physics}, 7, 2012.

\bibitem{SG16}
J~Poulter and J~A Blackman.
\newblock Properties of the ± j ising spin glass on the triangular lattice.
\newblock {\em Journal of Physics A: Mathematical and General}, 34(37):7527,
  2001.

\bibitem{EvenModel_SG1}
D.C. Mattis.
\newblock Solvable spin systems with random interactions.
\newblock {\em Physics Letters A}, 56(5):421 -- 422, 1976.

\bibitem{Theory_of_spin_glasses_P_W_Anderson}
S~F Edwards and P~W Anderson.
\newblock Theory of spin glasses.
\newblock {\em Journal of Physics F: Metal Physics}, 5(5):965, 1975.

\bibitem{SignedGraphs}
Thomas Zaslavsky.
\newblock Signed graphs.
\newblock {\em Discrete Applied Mathematics}, 4(1):47--74, 1982.

\bibitem{misfit}
S.~Kobe and K.~Handrich.
\newblock Correlation function and misfit in a computer-simulated
  two-dimensional amorphous ising antiferromagnet.
\newblock {\em physica status solidi (b)}, 73(1):K65--K67.

\bibitem{frustrationSQUpLowBonds}
Scott Kirkpatrick.
\newblock Frustration and ground-state degeneracy in spin glasses.
\newblock {\em Physical Review B}, 16(10):4630, 1977.

\bibitem{Wanglandau_Algorithm}
Fugao Wang and D.~P. Landau.
\newblock Efficient, multiple-range random walk algorithm to calculate the
  density of states.
\newblock {\em Phys. Rev. Lett.}, 86:2050--2053, Mar 2001.

\bibitem{EnergySizeEffect}
IA~Campbell, Alexander~K Hartmann, and Helmut~G Katzgraber.
\newblock Energy size effects of two-dimensional ising spin glasses.
\newblock {\em Physical Review B}, 70(5):054429, 2004.

\bibitem{PopulationDynamics}
M.~M{\'e}zard and G.~Parisi.
\newblock The bethe lattice spin glass revisited.
\newblock {\em The European Physical Journal B - Condensed Matter and Complex
  Systems}, 20(2):217--233, Mar 2001.

\bibitem{TcSqu}
H.~A. Kramers and G.~H. Wannier.
\newblock Statistics of the two-dimensional ferromagnet. part i.
\newblock {\em Phys. Rev.}, 60:252--262, Aug 1941.

\bibitem{TcTri}
R.~J. Baxter.
\newblock The inversion relation method for some two-dimensional exactly solved
  models in lattice statistics.
\newblock {\em Journal of Statistical Physics}, 28(1):1--41, May 1982.

\end{thebibliography}

\end{document}